# Deterministic and Nondeterministic Particle Motion with Interaction Mechanisms

**Cameron McNamee[1,2]\*, Renee Reijo Pera[1]**

[1]McLaugling Research Institute, Great Falls, MT, USA

[2]Department of Mathematics, California Institute of Technology, Pasadena, CA, USA

**\* Correspondence:**
cmcnamee@caltech.edu

**Abstract**

Studying systems where many individual bodies in motion interact with one another is a complex and interesting area. Simple mechanisms that may be determined for biological, chemical, or physical reasons can lead to astonishingly complex results that require a further understanding of the moving bodies. With the increasing interaction between computation and various scientific areas, it has become more useful, feasible, and important to create models for these systems. Here, we present two families of models, deterministic and nondeterministic, along with three distinct and realistic interaction mechanisms. These are combined in a unique way to provide the groundwork for particle system models across multiple disciplines. This work has applications that range from biology to chemistry and physics. In addition to describing the motivations and math behind all the models, software is provided that allows researchers to quickly adjust and implement what is described here.

## 1    Introduction

This work was inspired by consideration of quantum biology (QB) and cell migration and tissue formation as previously described **(Vecheck *et al.*, 2022)** with the goal of identifying quantitative measures of particle clustering in a scaffold. Identifying such measures involves observing the locations of the particles using image processing and applying these locations to a novel clustering algorithm. The results were some surprising numbers that led to a search for other, more robust methods of cluster analysis and an environment in which to test resulting algorithms. This proved not to be a simple task prompting development of potentially new models to understand how particles may interact under different conditions. From a biological perspective, and in view of the eventual application of this work, this paper explores the coalescing of particles in a space towards one another. It is also important to understand that in the simulation process, the goal is to replicate what is seen in realistic settings. In addition to biology, many systems wherein multiple autonomous particles are driven in a migratory pattern can be seen in physics and chemistry, allowing this work to be expanded to multiple fields **(Hansen *et al.*, 2013; E. Courtier *et al.*, 2019)**. Nanoparticle and ion motion are two distinct areas that this type of computational research can be readily applied and lead to interesting features of such a system being uncovered **(Hansen *et al.*, 2013; E. Courtier *et al.*, 2019)**. With the diverse opportunities in mind, this paper seeks to abstract the motion and interaction of individual bodies in such systems to create general models. These general models may then be taken by a researcher and applied with the appropriate parameters to match a particular system, with the addition of the available software laying the foundation of such a model **(Figure 1)**. This paper is by no means all-encompassing; however, the range of these general models spans many fields with many important implications.

## 2 Methods

All software was implemented in MATLAB. Methods are described below as follows: First, different environments in which particles may be generated are discussed. These environments decide what is called the "initial seeding" of the particles. These initial seedings vary in ways that may be more suitable for certain applications. Second, two families of models are explored: deterministic models and nondeterministic models. Differences between these two families of models are described in more detail; however, as the names suggest, the most important difference is that the deterministic model will produce the exact same results given the same initial seedings, while the nondeterministic models have an element of randomness that allows for changes in outcome. The third and last part discussed is the interactions between particles. This is a unique addition to these models and distinguishes this work for specific applications in various fields.

### 2.1 Initial Seeding Environments

Initial seeding environments are different environments that serve different purposes and are each suited for their own application. All seedings are centered at the origin.

### 2.1.1 Uniform Seeding

Non-random uniform seeding places each particle in a lattice point of a Cartesian coordinate system. This serves as what can be considered an "*in vitro*" environment, since in most applications, particles are not placed in such an orderly fashion (Peercy and Starz-Gaiano, 2020; Calvillo *et al.*, 2022).

### 2.1.2 Uniform Random Seeding

This seeding generates a designated number of particles within specified dimensions. This is carried out by generating three random numbers within $\pm$ the designated range for each particle. These coordinates then determine the placement of each particle, providing uniformly random particles throughout the space with an expected center at the origin.

### 2.1.3 Spherical Random Seeding

Like the seeding above, spherical random seeding is a uniform distribution of particles with an expected center at the origin. This differs from uniform random seeding in that the seeding is formed with a given number of particles and a radius. The problem of creating a uniform distribution within a sphere is not trivial; thus, this was accomplished by using polar coordinates.

To randomly generate a radius that places particles uniformly, a floating-point number between [0,1] is sampled and the cube root is taken. Then, the maximum radius is multiplied by the resulting value. This allows, for reasons that can be clearly seen from a geometric perspective, for the radius to generate uniform points. Generating the random angles is described in more detail here, along with more extensive reasoning throughout the entire process: (Simon, no date). This random process is done for as many particles as desired, and this determines the initial seeding.

### 2.2 Random Migration Models

This class of models stems from the Beauchemin model for migrating bodies, which has been used previously for biological modeling (Textor, Sinn and de Boer, 2013). All random migration models were taken from *Textor et al.* (Textor, Sinn and de Boer, 2013).

### 2.2.1 Beauchemin Random Walk

To develop a biased migration model for a group of particles, it is helpful to first create a model that has no bias. The Beauchemin random walk model functions under two phases: a pause phase and a free phase. During the pause phase, a particle generates two random angles, $\phi$ and $\theta$, and uses these angles to generate a uniform random direction. The generation of these angles follows the same method mentioned above for uniform sphere generation. After the pause phase, the particle enters the free phase. In this random walk model, every particle has the same, given velocity $v$. Using its given direction, a particle travels at the velocity $v$ for the entire free phase. The free phase length may be changed, but for most modeling applications, it is 4 times longer than the pause phase. This is arbitrary and may be changed easily.

As the name suggests, the Beauchemin random walk method is not biased in any way, so the particles walk uniformly in any direction. This unbiased nature leads to no pattern of particle migration but provides a basis wherein specific factors may be changed. With the unbiased model established, an introduction to different biases is now needed.

### 2.2.2 Simple Phenomenological

The Simple Phenomenological model (SP) introduces bias by adding a "drift element" during the pause phase (Textor, Sinn and de Boer, 2013). Instead of the particles remaining stationary during the pause phase, they drift towards some defined point. This causes uniform clumping amongst the particles. In this implementation, the choice was made to make this point the origin. This is of course arbitrary and may be changed to suit whatever environment is most sensible.

The drift velocity is defined as the given velocity $v$ times the bias variable $p$. For $p = 1$, equivalent bias is achieved between random walking and drifting towards the origin as the particles approach the center at the same rate that they walk randomly. Of course, for $p = 0$ there is no bias, so this model would operate the same as the random walk above.

The drift vector is defined by using the current $x, y,$ and $z$ coordinates to calculate the new coordinates. First, $d$, the distance from the drift particle, is calculated using the $l_2$ norm

$$d = \sqrt{x_0^2 + y_0^2 + z_0^2}$$

Then, the drift velocity is used to "shrink" the coordinates towards the origin

$$x_{drift} = x_0 \cdot \frac{d - vp}{d}$$

$$y_{drift} = y_0 \cdot \frac{d - vp}{d}$$

$$z_{drift} = z_0 \cdot \frac{d - vp}{d}$$

In this implementation, there is an added condition that a particle only moves towards the origin if it is farther away than the drift velocity will take it. This prevents the particle from "overshooting" its target.

### 2.2.3 Topotaxis Model

The Topotaxis Model (TM) differs from SP by introducing a different form of bias. Instead of moving the particle in the target direction, TM changes the distribution from which the angles are derived. Instead of $\phi$ and $\theta$ being taken from a uniform distribution, they are drawn from a beta distribution. A beta distribution is essentially a truncated normal distribution. The normal distribution cannot be used in this situation because of its infinite tails and the angles are within a strict range, $[0, 2\pi]$ and $[0, \pi]$ respectively. The beta distribution's shape is determined by two parameters. Here, the two parameters, $\alpha$ and $\beta$, and are set equal. Since they are equal, we only refer to $\alpha$ from here on. To calculate the shape of the distribution given a single bias variable $p$, we need to consider that when $p = 0$, the distribution should be uniform and as $p$ approaches 1, the distribution should become more skewed. This is accomplished by decreasing the variance in the distribution of angles while keeping the mean towards the origin with the result that the particles have an increased likelihood of orienting themselves towards the target location. In MATLAB, if $a = 1$, the distribution is defined as uniform and as $a \to \infty$, the variance decreases, and a peak centered at the mean value is formed. To calculate $a$ given $p$, the following is done

$$a = \frac{1}{1-p}$$

From this definition we see that for this model $p \neq 1$, but we also see $\lim_{p \to 1} a = \infty$, as desired. Thus, for $p = 0$, there is a uniform, or unbiased, distribution, and as $p \to 1$, there is an increased bias towards the targeted location.

## 2.3 Deterministic Migration Models

The deterministic migration models are a family of migration models that feature no random elements. Within this class of models, there are two different groups: uniform clustering and local clustering. The uniform clustering algorithms behave on a global scale, causing all particles to migrate in a wholistic sense. The local clustering algorithms organize themselves at a much smaller scale.

### 2.3.1 Uniform Clustering, the Naïve Approach

The first and simplest way of clustering particles within space is to "shrink" the space they are in. This first requires a clustering rate, $r \in [0,1]$. This method is as follows

$$U(c_x, c_y, c_z) = (c_x \cdot (1-r), c_y \cdot (1-r), c_z \cdot (1-r)); \forall c \in C \text{ (Eq. 1)}$$

Where $C$ is the set of all particles, $c$ is an individual particle, and $c_x, c_y$ and $c_z$ are the x-, y-, and z-coordinates of $c$, respectively. This function causes all particles to "march" inward, maintaining their relative distances to one another. This does produce a set where each particle is relatively closer to all other particles but does so in a way that is not representative of typical physical settings.

### 2.3.2 Nearest Neighbor Clustering

In nearest neighbor clustering, particles are not clustered into a single lump, but rather orient to their neighbors and move toward them. This approach is much more nuanced compared to Uniform Clustering and reveals samples that are much more reminiscent of certain applications. The method behind this approach assumes that particles can only sense their $n$ nearest neighbors (NN's), and so they only approach those particles. This method is conducted by first finding the geometric center (GC) described by the $n$ NN's and then moving the particle towards that at a clustering rate, $r \in [0,1]$. Using the same notation as above, we have the set of distances from all particles to a given particle $c$.

$$N_c = \min_n \{d_i : d_i = ||s_i - c||, \forall s_i \in C\} \text{ (Eq. 2)}$$

Here, $||s_i - c||$ is the $l_2$ normed distance between the particle $s_i$ and the particle $c$ and $\min_n(S)$ represents taking the $n$ smallest elements from the set $S$. With the NNs now identified, the GC ($g_c$) is calculated for a given $c$. This center is calculated without including $c$ itself.

$$g_{c_x} = \frac{\sum_{s \in N_c} s_x}{n}$$

$$g_{c_y} = \frac{\sum_{s \in N_c} s_y}{n}$$

$$g_{c_z} = \frac{\sum_{s \in N_c} s_z}{n}$$

$$g_c = \left(g_{c_x}, g_{c_y}, g_{c_z}\right) \text{ (Eq. 3)}$$

In other words, the unweighted mean of x-, y-, and z-coordinates of each NN. This gives us the GC described by the $n$ NN's. With this, we now have a location to move $c$ towards.

$$(c_x, c_y, c_z) \mapsto (c_x(1-r) + g_x r, c_y(1-r) + g_y r, c_z(1-r) + g_z r) \text{ (Eq. 4)}$$

For a given clustering, this calculation must be carried out for each particle, and it is done in its entirety before any particle moves (i.e., no particle's location changes until after every particle's new location has been calculated. Then, all particles move to their new locations at once). The result is clustering in small groups that after enough time steps, form individual groups.

### 2.3.3 Threshold Clustering

Threshold (TH) clustering differs from nearest neighbor clustering in that, instead of searching for a certain number of neighbors, the particles have an assumed maximum influence distance. This maximum influence distances, a threshold, is analogous to a particle only being able to sense other particles for a small radius around itself. This method is implemented by collecting the set of particles within the maximum sensing distance, $t$, calculating the GC described by these particles, and then moving towards this center at the clustering rate $r$.

$$T_c = \{s \in C : ||s - c|| < t\} \text{ (Eq. 5)}$$

Then, as above, the GC of this set is calculated using the unweighted mean (**Eq. 3**). This gives $g_c$, which we then use to calculate the new particle's position using **Eq. 4**. This, as with NN, is done for every particle before any particle changes its location. This mode produces several small groups throughout the space whose group size changes with $t$.

### 2.3.4 Weighted Clustering

This method is the next logical step for simulating clustering. Instead of taking the unweighted mean to calculate the GC, why not weigh the locations based on distance? Weighted clustering is done by first calculating $D_c$, the set of all distances from each particle to $c$. Next, the weighted geometric center (WGC) for $c$, $w_c$, is calculated as follows:

$$w_{c_x} = \frac{\sum_{i \in I_c} \frac{s_{i_x}}{d_i}}{\sum_{i \in I_c} \left(\frac{1}{d_i}\right)}$$

$$w_{c_y} = \frac{\sum_{i \in I_c} \frac{s_{i_y}}{d_i}}{\sum_{i \in I_c} \left(\frac{1}{d_i}\right)}$$

$$w_{c_z} = \frac{\sum_{i \in I_c} \frac{s_{i_z}}{d_i}}{\sum_{i \in I_c} \left(\frac{1}{d_i}\right)}$$

$$w_c = (w_{c_x}, w_{c_y}, w_{c_z}) \textbf{ (Eq. 6)}$$

Where $I_c$ is the indexing set corresponding to $D_c$. The reason for using the reciprocals of the weights follows from the idea that as a particle get further, it should affect $c$ less. The new coordinates for $c$ are calculated according to **Eq. 4**. This is a global clustering on particles, like uniform clustering. This causes no small groups to small, but rather the entire population of particles congregate. This congregation, however, does not appear the same nor does it occur at the same rate at uniform clustering.

### 2.3.5 Weighted Nearest Neighbor Clustering

In accordance with the previous modes, this weighted nearest neighbor (WNN) clustering is the next intuitive step. This method is, as the name suggests, a combination of NN clustering and weighted clustering. This combines the logic of only looking towards the closest particles, while also weighing the particles according to their distance. The set of NN's is calculated according to **Eq. 2**.

$$I_{N_C} = \{n \in I : s_n \in N_c\}$$

The WGC is calculated according to **Eq. 6** with $I_{N_C}$ used in place of $I_c$. With this WGC, particle $c$ is mapped using **Eq. 4**. This mode appears very similar to the original NN clustering but tends to form clusters that are skewed in slightly different locations.

### 2.3.6 Weighted Threshold Clustering

Weighted threshold (WT) clustering follows the same intuition. $T_c$ is calculated according to **Eq. 5.** This defines the limited distance indexing set:

$$I_{T_c} = \{n \in I : s_n \in T_c\}$$

Using **Eq. 6** with $I_{T_c}$ in place of $I_c$ gives WGC about the particles within the threshold. **Eq. 4** is used to map $c$. As with WNN clustering, WT clustering appears quite similar to the TH clustering on small scales.

**2.4 Interaction Mechanics**

With the addition of interaction, defined here as how two particles behave when they physically hit each other in the migration process, there are many different fields that may provide insights. Some outcomes of physical interaction are cessation of movement, repolarization, or adhesion, to name just a few examples of how particles may interact with one another.

**2.4.1 Cessation of Movement**

The simplest self-interaction is cessation of movement, meaning simply that if two particles are within a designated distance of one another, they both stop moving. In simulation, implementation of this outcome simply checks if a particle is close to any other particle. If it is, it stays in the same spot. Otherwise, it keeps moving as it was intended.

**2.4.2 Repolarization**

Repolarization means that if two particles hit each other, they both turn around in the opposite direction, much like an elastic collision. Implementation is carried out by taking the two angles that defined the direction that a given particle was traveling, and negating both.

**2.4.3 Adhesion**

Adhesion is the clustering or clumping of particles to form a single body that moves as one. This is accomplished by assigning a clump to every particle. If a particle is within a given radius of other particles, it gets put into that clump. Additionally, a clump is a collection of not just the particles that are touching a given particle, but any that are touching its neighbors, the neighbor's neighbors, and so on. The clumps partition the set of particles and create moving bodies that can obtain more particles. No limit to the size of clumps was designated, however this addition is trivial.

**3 Discussion**

The purpose of this discussion is to explore the potential applications of the models described above and how the simulations generated in the linked code can prove useful to other researchers. Many particle or particle-like systems are seen across all scientific fields including physics, chemistry, and biology. Proper implementation of these models requires a good understanding of the system at hand and how the particles behave with one another. An understanding of the system is necessary to select the proper migration model.

In highly chaotic systems, nondeterministic models are more suited as randomness (pseudo randomness rather) is a key feature of such a system. In systems where the environment itself does not drive motion of particles themselves, the deterministic models are better approximations.

Implementing different interaction dynamics in the same setup leads to very different results. Thus, it is also very important to understand how individual particles behave upon collision. With this necessity understood and considered, let us explore specific use-cases.

The above models offer insight in biological settings. Cell migration is an important component of many cell types, and thus estimating that motion can provide information about cellular mechanisms (Davis *et al.*, 2015; Gopinathan and Gov, 2019; Metzcar *et al.*, 2019; Norden and Lecaudey, 2019; Alert and Trepat, 2020; Guberman, Sherief and Regan, 2020; Peercy and Starz-Gaiano, 2020; SenGupta, Parent and Bear, 2021). In different settings, cellular motion may be estimated by different models. Chemotaxis, which is motion based on the sensing of chemical gradients, is the driving force of many migration patterns at the cellular level (Hu *et al.*, 2010; Gopinathan and Gov, 2019; Norden and Lecaudey, 2019; Alert and Trepat, 2020; SenGupta, Parent and Bear, 2021). This mode of migration is akin to the TM as the chemical gradient introduces bias as to what direction the cells will migrate. The stronger the cells' reaction to the chemical gradient, or the stronger the chemical gradient itself, the greater this bias. Migratory cells have been noted to have a degree of continuous bias, meaning they can have varying levels of directional influence due to chemotaxis, not simply a binary model of directed motion (Hu *et al.*, 2010; Alert and Trepat, 2020). This, along with the mechanistic understanding of chemotaxis lends this type of cellular motion to the Topotaxis model.

In addition to chemical sensitivity, further support of the nondeterministic models comes from the observations that cells perform motion that follow the pause phase-free phase mechanism that dictates these models (Gopinathan and Gov, 2019; Alert and Trepat, 2020; Peercy and Starz-Gaiano, 2020). This comes from the discontinuation of motion in favor of direction recalibration noted in cell motion, which strongly motivates the use of Beauchemin random walks in favor of other forms of random walks (Textor, Sinn and de Boer, 2013; Gopinathan and Gov, 2019; Alert and Trepat, 2020; Peercy and Starz-Gaiano, 2020). Other than chemical gradients, the mechanical force induced by the microenvironment about the cell can influence motion (Peercy and Starz-Gaiano, 2020; SenGupta, Parent and Bear, 2021). This is the defining feature of the SP model as there is an external force on the cells driving them in a biased direction, much like fluid flowing. SP can be used to describe mechanistic motion of cells during embryonic development and within the neural crest (Giniūnaitė *et al.*, 2020; Peercy and Starz-Gaiano, 2020). Contexts such as these are strong candidates for applying nondeterministic models outlined in this paper in the field on biology.

Deterministic models also have niches in biological settings. Tumor growth has many similarities to TH, as does cancer cell migration (Metzcar *et al.*, 2019). This is an especially important area for computational research, as hypotheses must be tested efficiently and accurately. Modeling can also be used to determine metabolism of cancer and stem cells, giving useful information that can be verified experimentally (Metzcar *et al.*, 2019). This is a great example of how computational modeling can uncover additional information and provide meaningful insight in a timely fashion.

Cellular mechanics also offer interesting interaction dynamics. All three modes of interaction take place across the field of biology. Adhesion, as described above, occurs in migrating cell groups, cancer metastasis, and embryogenesis (Gopinathan and Gov, 2019; Norden and Lecaudey, 2019; Alert and Trepat, 2020; Guberman, Sherief and Regan, 2020; Peercy and Starz-Gaiano, 2020). This typically occurs in homogenous cell groupings and is seen throughout the immune system (Davis *et al.*, 2015; Norden and Lecaudey, 2019; Alert and Trepat, 2020). Cessation of movement can be observed in tissue growth for both epithelial cells and fibroblast cells (Abercrombie, 1961; Guberman, Sherief and Regan, 2020). In fact, this behavior is the original mechanism that led to the term "contact inhibition of locomotion," better known as CIL (Abercrombie, 1961). Notably, CIL also can be described in different contexts as a repolarization effect, meaning within biology any of the above interaction mechanisms may be observed (Davis *et al.*, 2015; Roycroft *et al.*, 2018; Alert and Trepat, 2020; Guberman, Sherief and Regan, 2020).

Moving to a different field, in chemistry and physics nanoparticles display many of the features of both clustering and motion. In certain settings, nanoparticles can behave in random walk migration which can be biased due to outside forces to appear as any of the nondeterministic models outlined above (Hansen *et al.*, 2013). In addition, a mode of random potential driven motion along with particle adhesion described as Oswald Ripening can be appropriately described by the TM with particle adhesion (Hansen *et al.*, 2013). Ion motion in solar cells holds many of the key mechanistic contexts outlined above (E. Courtier *et al.*, 2019). Additionally, the dynamic change of chemical gradients, useful for chemistry and biological contexts, can by approximated by SP (Hu *et al.*, 2010).

The novelty of the work contained within this paper comes from the diverse applications readily available. It is the addition of particle interactions that leads to results with astounding complexity that are reminiscent of what is captured in scientific settings, while keeping the simulations in a tractable form. While this work was inspired from a biological setting, it is of upmost importance to recognize the applications of such work to other fields, for it is the interaction between the sciences that leads to strides in innovation. The union between computer science and physical science is something that has never been seen before, and we as a scientific community have but scratched the surface.

With these diverse and important applications available from the relatively simple models outlined above, it is important to note the limitations of not just these models themselves, but also the current state of computational physical science. First, we can only model what we understand, and this knowledge is limited not by our ability to create models but by our methods in the physical sciences. To aptly describe a complex system, a complex understanding of such system must be had first. However, as far as biology, much of chemistry, and many places of physics goes, we do not have such a deep understanding (SenGupta, Parent and Bear, 2021). The power of predictive models is the ability to extrapolate our current understanding to generate nuanced hypotheses that may be tested in the future. A major pitfall of this ability is, however, that false initial assumptions may lead to a fruitless path.

The goal and purpose of this newfound computational power, a power that is only growing as time passes, is to approximate to the best of our ability the real world. This allows us to uncover, with unparalleled speed, extremely important new aspects of physical systems that were previously obscure and serve to inspire experimentalists. Continuing to cautiously push forward with this technology will lead to profound discoveries across academia.

*The authors declare that the research was conducted in the absence of any commercial or financial relationships that could be construed as a potential conflict of interest.*

CM contributed to the experimental design, analysis of data and interpretation of the results. CM and RRP prepared the manuscript and edited the final manuscript for publication.

This work was supported by funding from the National Institute of Health (1R01HD096026).

## 9 Data Availability Statement

The software for this study can be found at "Cameron-McNamee/Particle_Modeling" [https://github.com/Cameron-McNamee/Particle_Modeling].